\documentstyle[spie_a4]{article}

\begin{document}

\title{Operational Concepts of Large Telescopes\footnote{~To
appear in the Proceedings of the SPIE Conference
``Optical Telescopes of Today and Tomorrow,'' A. Arneberg (ed.), held
May 29 - June 2, 1996, in Landskrona/Hven, Sweden}}

\author{Bruno Leibundgut\\[12pt]
        European Southern Observatory\\
        Karl-Schwarzschild-Strasse 2 \\
        D-85748 Garching \\
        Germany}

\maketitle

\abstract{
The singular way of scheduling at major
observatories has favored certain types of astronomy. 
This has led to the discrimination of certain types of observations
which could not be accommodated. 
It is the goal of future operations to open
possibilities for some additional types of observational astronomy. 
Together with
improved observatory operations the new observing modes can provide
significant progress in the acquisition of astronomical data.
Specifically, the capability to optimally match the schedule to observations
should prove a major advantage of service observing. 

The VLT data flow project is designed to accommodate these new observing modes
together with conventional observing.
Simulations and first experience with the ESO NTT will prepare and
refine the concepts and procedures foreseen for the VLT. The
acceptance of service observing by the astronomical community is of 
critical importance for this new operational mode to succeed.  }


\rm

\section{INTRODUCTION}

The efficiency of the new large telescopes to increase our
knowledge of the universe and its constituents will depend
significantly on how astronomers can utilize them. 
Only creatively-used and well-operated telescopes
will be able to contribute their share to the development of
astronomy. 

Operational concepts currently discussed by the various large
telescope projects can be divided into two main groups which differ
mainly by the composition of their user communities. On the one side are
the privately-owned observatories with access restricted to a small
community of astronomers. For many of these observatories it
is most economical and also easiest to continue operation of their
telescopes as in the past. The astronomers
perform the observations themselves and are fully responsible for
the acquired data. Normally they are assisted by experienced
telescope operators. In the following this situation will be referred
to as conventional observing, often also referred to as
classical observing. 

National and international observatories on the
other hand are actively experimenting with service, or queue,
observing modes$^{1,2,3}$. The attempt is to take better advantage of the
varying conditions by combining the best-suited observations independent
of individual astronomical projects (cf. [4]). 
Active interaction with the scientist,
who proposed the observations and will analyze the data, during the observing
process becomes
nearly impossible in this case. The break in the
observational chain has to be recovered by operational procedures
which enable the astronomers to maintain control over their
observations. At the same time interesting new
astronomical possibilities emerge with such a mode. 

The next
generation instrumentation which comes with the new telescopes will
also deliver unprecedented data at the cost of increased complexity.
Astronomers will need every assistance in preparation and execution of
their observations possible. Often when observations are defined in 
sufficient detail the actual data acquisition can be delegated to the 
observatory. Many aspects of the advantages of the various modes and
the changes involved have been collected in [4].

In the following we will discuss some of the advantages of the two
observational modes and develop criteria when they are suited best (\S2).
The concepts of the VLT specific data flow project which is designed 
to accommodate the needs of service observing are presented in section 3. 
Open issues and future directions are discussed in the conclusions.

\section{ADVANTAGES --- ONE WAY OR ANOTHER}

The way current observatories are operated has been developed to
efficiently distribute the scarce resource of observing time. 
Major guidelines
are scientific merit of the proposed observations, fair
distribution among the subfields of astronomy and -- to a certain
degree -- democracy. Each successful applicant
is awarded a certain amount of time at fixed
calendar dates to perform the experiment. All the observatory provides
is a functioning telescope and instruments with no or little
guidance on how to best perform the observations. The observatory's 
r\^ole is essentially to administer the resources while developing
and maintaining the infrastructure. 

With the astronomers guiding the observations at the
telescope a quick scientific assessment of the data in respect to their
suitability for the project can be done. The vital link of the
astronomical observing experience and the researcher is maintained
which assures that the data can be analyzed properly. Psychological
aspects like the personal involvement of the astronomers with the data
acquisition or the astronomers' detachment from their
regular work during observing trips can also contribute to the
creativity of the scientific process. Advantages of this conventional 
mode for the observatory are the interaction of observatory staff
with the astronomers, the experience brought in by the external users
to the observing process, and the possibility to transfer the
responsibility of the data acquisition to the astronomers directly.
This is particularly important in cases of specialized observations,
the outcome of which can not be predicted. 

Typical research favored by these operations is based on observations
which can be obtained under ``regular'' environmental conditions 
delivered
by the site and the telescope. Projects which require special
conditions are clearly discriminated, unless there is easy
access to the facilities as is often the case at 
observatories with small user communities. There, astronomers 
typically obtain a larger share of observing time and can
mix their own projects to make adequate use of particular
meteorological circumstances. Other projects which are typically
disfavored in this mode are surveys since the normally large time
demands can not be readily allocated at observatories which serve
large communities. Another type of project which normally suffers
from this conventional scheduling mode are targets of opportunity
and general time series observations of objects with time scales
of weeks or months. Smaller communities, where the data exchange is
organized informally, have been able to arrange for such occasions, but
the larger observatories had to introduce formal and often
awkward rules to handle these situations. 

The previous paragraph lists a few astronomical reasons why it would
be advantageous to change scheduling procedures. The exploration of
parameter space mostly inaccessible so far can add
significantly to astronomical progress. It is this widening of the
observational options which is scientifically most interesting and
leads to the discussion of service observing.
The possibility
to select observations matching the prevailing conditions best will
enable projects depending on special circumstances.
This observational edge may speed up results
which otherwise would rely on the meteorological luck of the draw. 
Surveys can be
carried out mixed with other observations. It has to be noted
that most massive surveys have been obtained in service mode as
the example of the Palomar Observatory and the ESO/SERC Sky Surveys, 
the 
current near-IR surveys (DENIS and 2MASS), or the searches for massive
halo objects (MACHO, EROS, OGLE) with their scientific spin-offs show. 
Another observational niche difficult to access at large
observatories is the monitoring of variable objects and observations
of targets of opportunity. Such synoptic observations can very easily
be fitted into a flexible scheduling process, which is a prerequisite 
for service observing. A small number of objects spaced around the 
sky requiring just a few observations can easily be accommodated in 
service mode, while they may not reach critical program size for 
time allocation in the conventional case.
Another aspect more difficult to quantify may be the
more efficient use of the available time as programs will be performed
to the exact amount of exposures needed. Finer adjustment to the 
moon phases gains some dark hours which are often lost when programs 
are scheduled conventionally in blocks of complete nights$^{5,6}$.
Observing projects may improve by the needed preparations which entail a
complete road maps of how the data will be analyzed. This should
happen even though there will be no formal requirement to do so. 
A further 
possible spinoff is the availability of an extensive archive of
observations. Archival research, although not yet a primary resource,
may contribute significantly to projects where the combination
of data from different wavelength regimes is advantageous.

While there remain many astronomical projects which clearly are
best served in conventional mode, the above examples show that
there are a few reasons why it may be interesting to explore some
other operational modes with new telescopes. The selection of
the most appropriate observing mode for each project will be an 
important decision which will need careful evaluation by the
astronomer.

The move to new operational schemes must be driven
by improvements in the scientific process. Many discussions on
operational issues have focussed on advantages for observatories, 
which are not negligible, but can not by themselves justify the
proposed, stringent changes. The astronomical community must
embrace the new opportunities for success. It should also be noted
that all observatories plan to offer both observing modes 
leaving all options open.

\section{INFORMATION EXCHANGE IN A COMPLEX OBSERVATORY ~~~~~~~~ THE VLT DATA FLOW}

Removing the astronomers from the actual observations at the telescope 
implies that
other means must be provided to retain their control over the 
observational process. The development of the
procedures to guarantee the astronomer's participation has started at
ESO within the On-Line Data Flow project (see also [1]).
Its basis was developed in the VLT
Science Operations Plan$^5$ and a document which defines the
astronomical requirements on the observatory information chain$^7$.
The link between the astronomer and the observatory should be close
and transparent. It also should remain flexible.

There are a few very clearly separated stages in the astronomical 
information and data cycle$^{1,7}$.
Each phase has demands and services which have
to be identified and carefully combined. The definition of
the observing program and its scientific evaluation by peers as well as the
definition of the individual observations should be provided by the
astronomy community. The observatory solely supports and administers 
this process. Scheduling of programs in conventional mode and
the observations in service mode, however, are performed by 
the observatory. 
The demand for exact observation definition drives the
requirement for the second phase of the astronomer -- observatory 
interaction.
To built a schedule which optimally matches the prevailing conditions
with the observational requirements the scheduler will need accurate
input from a meteorological site monitor, telescope and instrument
status, and astronomical restrictions (e.g. moon phase). The actual
observations will be handled by specific telescope, instrument, and
detector software$^8$ which will return the
data products (frames and logs) to the data flow. The further data 
handling involves
archiving and pipeline processing for provisional quality control. All
these processes fall into the responsibility of the observatory.

These considerations led to the adoption of a two-layered system
relaying information among the various subprocesses. The VLT could be
run from the control system alone without the data flow
superstructure, but the technical description of the
instrument and the interaction with the scheduling process are
considered too detailed and cumbersome for astronomers who infrequently 
interact with the system. The data flow acts as the intermediary between the 
astronomers and the technical software. There are four fundamental
agents the data flow is connecting: the astronomer, the scheduler,
the technical software, i.e. the observing facilities, and the archive. 
The VLT concept does reflect their basic needs. ``Observation
Blocks'' contain the complete information relevant for an
individual observation$^{1,9}$.
An observation in what follows is
considered a single pointing of the telescope with a specific instrument 
setup to acquire a coherent data set.
Apart from obvious quantities, like
coordinates and instrumental setup, observation blocks can also
contain global requirements of importance to the scheduler, general
comments of interest to the observer, and links to reduction
procedures or quality control. A specific feature is the modularity by
which observation blocks can refer to other observation blocks to
combine observations. It is thus possible, e.g., to link regular
observations with the acquisition of calibration data.

All information on the instrument and its operation during the
observation is encapsulated in ``instrument templates$^{1,9}$.'' These
structures define commonly used setups and are embedded in the 
observation blocks. The astronomer defines the
specific parameters of the setup in a template parameter file 
which accompanies
the template. This should ease the astronomers' interaction with the
VLT system as many details of instrument operations can be served by
the templates. Some observations will not be offered with templates in
which case the option to
drop to the level of the VLT technical software is still available.

Astronomers whose proposals successfully passed the selection process
will hence have to prepare the observation blocks and the templates for
each observation in their program. This preparatory phase 
is foreseen for all proposals and guarantees the close
involvement of the astronomers. Even conventional observations
will be prepared in this preparatory phase to familiarize the
astronomers with the system. In this case the astronomers will use
their observation blocks at the telescope during their assigned
nights. During service time all available observation blocks are
provided in a central database polled by the scheduling software. The
scheduling is a very complex process which currently is not yet fully
defined for the VLT. Since the best performance is achieved only when
sufficient information is available and the detailed procedure depends
on many different sources, it is important to collect as much 
intelligence as possible. Observations with detailed descriptions 
of their requirements are more likely to achieve the requested 
quality as they can be scheduled accordingly.

A long-term plan for the semester or a fair fraction of it schedules 
conventional observing runs and defines the requested instrument 
setups. This will be required even with instrument changes becoming 
possible at short notice as special filters or gratings will have 
to be mounted ahead of time. 
Flexible scheduling itself will rely on local information sources 
which describe the prevailing conditions of and at the observatory. 
Meteorological input is provided by a site monitor.
Image quality assessment (including 
sky background), possible forecast of critical parameters (e.g. cloud 
cover, precipitable water vapor, seeing) will provide the basis of 
the selection process. 
ESO has maintained a program to characterize prevailing
meteorological conditions over the past several years$^{10}$
and is embarking on a
project to forecast some of the meteorological parameters on the basis 
of a few hours. 
Options for operations with limited information have to be developed
as well. 

An important aspect of the scheduling process is the underlying
criteria which govern the selection. This is largely unexplored
territory for all observatories with the notable exceptions of HST and
the NOAO operations of the WYIN telescope$^2$. Important results are
expected from simulations with mock projects which encompass a large
set of observations and a variety of requirements. Although it will
be impossible to fully simulate the scheduling of a semester
describing
the creativity of programs of real astronomers and vagaries of
real-time operations, simple strategies can be tested and compared. The
effect of operational overheads, instrument changes, 
decision time scales, 
importance of program completion, and weighing of different
conditions can be explored ahead of time. 

The telescope and instrument software returns raw data frames to the
data flow. Archiving and further processing complete the cycle. The
data archive captures all relevant information for a given
observation. This includes the original request contained in the 
observation block and template together with the actual 
conditions. Other relevant observations, e.g. calibrations and 
standard star data, linked to the project are also 
stored in this central place. The astronomers will receive 
(or retrieve) 
their data from this archive. Once data become public it will be
accessible by the whole astronomical community.

At the VLT a routine pipeline processing of all data obtained in
service mode and, possibly, conventional observing will
be attempted. The pipeline results are used for a quick assessment
of the data, potentially influencing the further observations of the
night. They will provide preliminary removal of instrumental and detector 
effects.
The quality control will follow the pipeline reductions to ensure that
the observations correspond to what was requested.

To test the concepts and procedures of the data flow a set of
reference proposals has been defined. These observational projects
with some scientific background have been designed to cover a large
range of observational requests and techniques. They will be used to
check the interfaces and the interactions of the various parts of the
data flow. For a first check they will play the r\^ole of external
astronomers. The reference proposals can also be used for simple 
scheduling simulations. 

\section{BUILDING THE EXPERIENCE}

The complexity of the operations of modern large telescopes should
not be underestimated. The required information exchange between the
astronomers and the observatory represents a vital link to assure
that the observatory delivers what is requested and expected by the
astronomers. The future will look different for the regular user 
even observing conventionally. Telescope operations
have been long ago delegated to specialists and astronomers have
accepted the help provided by telescope operators. The complexity
of the instrumentation and the observational procedures will further
emphasize the astronomers' understanding of technical aspects.
It should be the goal of the operations to keep the astronomers' 
interaction with the facilities and the staff as simple as possible. 
Every
possible help the observatory can provide should be available to
support the astronomers in their scientific experiments. They should
be able to concentrate on the observational aspects rather than
technicalities. Nevertheless, the astronomers will have
to be provided with sufficient information so that they can understand
and assess their data in all observational aspects. 

To assure acceptance of these conceptual changes the
collaboration of the astronomical community has to be assured.
The early involvement of future users of the system
can only improve the operations. Several science test cases for the
VLT have been solicited from the European astronomical community.
These test cases will be used in addition to the internal reference
proposals to test the procedures. They have the additional advantage
to be based on real science projects and the external astronomers are
experts in the requested observations who can provide helpful
criticism. Since the data flow is the VLT's
interface with the astronomical community it will largely
define the perception of the observatory. The input from users is
essential for a successful development.

The recommissioning of ESO's New Technology Telescope (NTT) will be
combined with the start of service observing at ESO. Several programs
have been approved and the data flow system will undergo a first real
test before the end of 1996. At first, the service mode will be restricted to
direct imaging in the optical, some of the operationally least
demanding observations. Spectroscopic observations in the service 
mode will offered only during the following semester.

\section{CONCLUSIONS}

The introduction of new observing modes in combination with the
improved instrumental capabilities is a daunting task. The prospective
advantages are significant and may provide an important observational
edge over other approaches. At the same time the success of the
experiment almost entirely depends on the acceptance by the community.
The astronomers will have to learn to optimally use the new
possibilities. Since many of the changes have been initiated by the
observatories it will be their r\^ole to convince the rest of the
community of the gains. A fundamental requirement is the smooth
operation of the observatory and the improved data quality has to
become an essential argument.

At the VLT the data flow will link the observatory to the astronomers
and expand the interaction between them. It presents the astronomers
with all observational possibilities and lets them make best
use of the facilities. An integrated approach has yielded a
system which will entail all operational aspects. It should be noted
that there exists a clear separation between the needed
infrastructure, the data flow, and the operational model. A flexible
data flow system will provide the options to built and improve
operational models without major limitations. 

Definition of an operational model for the VLT will have to tackle
open questions like the scale of the preparatory phase, the
exact criteria for the scheduling process, and 
the degree of pipeline processing. A convincing model must include
compelling reasons for the expanded preparatory phase. Flexible
scheduling drives most of the complexity of the VLT data cycle. The
operational model will have to set the astronomical priorities,
possibly based on the results from simulations.

First lessons from
simulating the information and data cycle with mock projects will be
followed by service operations of the NTT in early 1997. Further
refinement in the VLT environment at the NTT$^{11}$
will provide a solid basis for a successful start of service 
observations at the VLT itself.

Service observing will remain an experiment for the first few years.
It must not be seen in isolation as it is introduced to compliment the
current observational capabilities. Observing with the VLT will be
possible in conventional as well as service modes.
The selection of which mode suits the program and its observation best
should ideally be based on astronomical criteria. This can be achieved
when sufficient trust has been built through reliable delivery of
high-quality data.

\section{ACKNOWLEDGEMENTS}
Building the operations of a complete observatory is not a small
task and depends on many people. The views expressed in this article 
are based on discussions with many colleagues at ESO. The data flow
project is headed by P. Quinn and includes M. Albrecht, E. Allaert, 
D. Baade, A. M. Chavan, P. Grosb\o l, M. Peron, G. Raffi, and 
J. Spyromilio. They have been instrumental in developing some of
the ideas. I am also grateful to A. Renzini, the VLT project scientist,
for many discussions regarding these issues.


\begin{thebibliography}{99}

\bibitem{}{}{} Peron, M. \& Grosb\o l, P. 1996, these proceedings 
\bibitem{}{}{} Silva, D. 1996, NOAO Newsletter 47   
\bibitem{}{}{} Puxley, P. et al. 1996, these proceedings 
\bibitem{}{}{} Davies, J., Robson, I., \& Boroson, T. (eds.) 1996, {\it New
Observing Modes for the Next Century}, ASP Conference Series 87 
\bibitem{}{}{} Baade, D. 1995, ``VLT Science Operations Plan'', 
VLT-SPE-ESO-10000-0441, Garching: ESO 
\bibitem{}{}{} Boroson, T. 1996, in {\it New Observing Modes for the Next
Century}, eds. J. Davies, I. Robson, and T. Boroson, San Francisco:
ASP Conference Series 
\bibitem{}{}{} Grosb\o l, P. \& Leibundgut, B. 1995, 
``VLT On-Line Data Flow: Requirement Specification'', 
VLT-SPE-ESO-10100-0749, Garching: ESO 
\bibitem{}{}{} Wirenstrand, K. \& Raffi, G. 1996, these proceedings 
\bibitem{}{}{} Leibundgut B. 1996, in {\it New Observing Modes for the Next
Century}, eds. J. Davies, I. Robson, and T. Boroson, San Francisco:
ASP Conference Series 
\bibitem{}{}{} Sarazin, M. 1996, in {\it New Observing Modes for the Next
Century}, eds. J. Davies, I. Robson, and T. Boroson, San Francisco:
ASP Conference Series 
\bibitem{}{}{}Wallander, A. 1996, these proceedings 
\end{thebibliography}
\end{document}